\documentclass{jfm}

\usepackage{graphicx}
\usepackage{epstopdf,epsfig}
\usepackage{amsmath,amssymb,bm,mathrsfs}
\usepackage{booktabs}
\usepackage{xcolor}

\newcommand{\dd}{\mathrm{d}}
\newcommand{\avg}[1]{\left\langle #1\right\rangle}
\newcommand{\havg}[1]{\overline{#1}}
\newcommand{\uu}{\bm{u}'}
\newcommand{\ue}{\widetilde{\bm{u}}}
\newcommand{\we}{\widetilde{w}}
\newcommand{\safeincludegraphics}[2][]{%
  \IfFileExists{#2}{\includegraphics[#1]{#2}}{%
    \parbox[c][0.22\textheight][c]{0.92\linewidth}{\centering
    Figure file not found: \texttt{\detokenize{#2}}}%
  }%
}

\shorttitle{Marginal Energy Stability Theory}
\shortauthor{Z. Ding, B. Wen \& H. Li}

\title{Optimal heat transport at the edge of energy stability}

\author[Z. Ding, B. Wen \& H. Li]{Zijing Ding$^{1,2}$\corresp{\email{z.ding@hit.edu.cn}}
\ns\&\ns Baole Wen$^{3}$
\ns\&\ns Hui Li$^{1,2}$}

\affiliation{$^{1}$International Research Center for Intelligent Fluid Mechanics, Harbin Institute of Technology, Harbin 150001, China\\
$^{2}$School of Civil Engineering, Harbin Institute of Technology, Harbin 150001, China\\
$^{3}$Department of Mathematics, New York Institute of Technology, Old Westbury, NY 11568, USA}

\date{?; revised ?; accepted ?.}

\begin{document}

\maketitle

\begin{abstract}

High heat transfer in thermal convection is commonly associated with vigorous turbulent motion, but turbulence intensity alone does not explain how a limiting transport state is selected. We propose that such limiting states are organized by marginal energy stability. Starting from the exact perturbation-energy balance, we determine, for a prescribed mean temperature profile, the smallest neutral Rayleigh number over the balance parameter and all admissible disturbances. The corresponding marginal modes are then coupled to the exact mean-temperature equation, producing a self-consistent profile and heat flux. At large $Ra$, the selected branch gives $Nu\simeq0.0245Ra^{1/2}$ and develops conductive inner layers, logarithmic-like intermediate layers, and a weakly stably stratified core. An equivalent backg  round Lagrangian formulation yields the same governing equations and establishes uniqueness of the selected mean profile. Three-dimensional simulations at $10^6\le Ra\le10^8$ confirm that prescribed thermal forcing based on this profile drives the system toward a quiescent conductive state for Rayleigh numbers at or below the design value $Ra_c$, while the energy-stability condition is violated and residual convection persists for $Ra>Ra_c$. More broadly, these results establish marginal energy stability as a predictive framework for controlling transport far from equilibrium, offering a design principle for convection-free thermal management in systems where fluid motion is traditionally detrimental.

\end{abstract}

\begin{keywords}
Rayleigh--B\'enard convection, energy stability, flow control
\end{keywords}

\section{Introduction}\label{sec:introduction}

Thermal convection is a canonical example of nonequilibrium transport, governing heat exchange in planetary interiors, stellar envelopes, geophysical flows, industrial cooling, and materials processing \citep{Ahlers2009,Chilla2012}. In Rayleigh--B\'enard convection, a fluid layer heated from below and cooled from above becomes unstable when buoyancy overcomes viscous and thermal diffusion.
Above onset, fluid motion enhances the vertical heat flux, and the dimensionless heat transport, measured by the Nusselt number $Nu$, typically increases with the Rayleigh number $Ra$. At large $Ra$, convection becomes turbulent, and heat transport exhibits power-law scalings in $Ra$. Although experiments and direct numerical simulations (DNS) have reported a range of effective scaling exponents and regime transitions across accessible values of $Ra$~\citep{Chavanne97,Niemela,Ahlers2009,Urban2011,Xia2013,Iyer,Lepot,Bouillaut,Jiang,Xia2025}, the exact asymptotic form of this transport scaling remains unsettled. Rather than fitting empirical exponents to experimental or DNS data, we address more fundamental questions: what is the maximal heat flux achievable by convection, and what physical principles select the flow structures that approach this upper bound?

One approach to these questions is to seek exact coherent solutions to the full governing equations that maximize heat transport. Recent studies demonstrate that exact steady solutions can capture many features of turbulent convection observed in time-dependent simulations \citep{Waleffe2015,Sondak2015,Kooloth2021,Wen2020JFM,Wen2022JFM,Motoki2021,He2026,Ding2021,Ouyang2025}. Moreover, certain exact steady solutions---when optimized for $Nu$-maximizing aspect ratios---transport significantly more heat than turbulent convection at comparable parameters~\citep{Wen2020JFM,Wen2022JFM,Ding2021,Ouyang2025,He2026}. Notably, these optimal states exhibit the classical asymptotic scaling $Nu \sim O(Ra^{1/3})$, consistent with Malkus's marginal linear stability theory~\citep{Malkus1954,Howard1963}, which posits that convective heat transport is maximized subject to the constraint that the thermal boundary layer remains marginally stable to linear perturbations. To date, this $1/3$ exponent represents the highest scaling identified among exact coherent solutions in convection.

Whether a higher scaling is physically attainable remains an open question. Rigorous upper-bound analysis \citep{Busse1969,Howard1972,DC1996,Plasting2003} yields $Nu\lesssim Ra^{1/2}$ for convection between no-slip boundaries at finite Prandtl numbers, suggesting that the $1/3$ exponent of exact coherent solutions may not represent the true ceiling. The background method, one of the central tools in this framework, bounds the heat transport by combining an auxiliary temperature profile with an energy-like inequality that must hold for every allowable disturbance \citep{DC1996,DingKerswell20}. Within this framework, however, only the continuity equation (incompressibility) is strictly enforced as a physical constraint. Because the auxiliary temperature profile is not required to satisfy the full physical heat equation, the upper-bound analysis is largely decoupled from the actual flow dynamics, and the extremal states associated with rigorous bounds typically lack a direct physical interpretation. Rigorous bounds constrain what is possible, but they do not by themselves identify which physical mechanisms select the extremal configurations or whether the $Ra^{1/2}$ scaling is dynamically realizable.

Wall-to-wall optimal transport provides a complementary perspective on this question \citep{Hassanzadeh2014,Tobasco2017,Alben2017,Motoki2018,Souza2020,Doering2019}. By optimizing over incompressible velocity fields at a prescribed flow intensity, this method produces highly efficient, organized flows whose heat transport approaches rigorous bounds. The wall-to-wall approach incorporates more physical information than standard upper-bound theory: it enforces both the full heat equation and incompressibility as constraints, even though the resulting velocity fields may not satisfy the momentum equation. However, these optimal flows can theoretically be realized by introducing a tailored body force into the momentum equation. Between no-slip boundaries, these two- and three-dimensional flows exhibit hierarchical, self-similar structures and achieve a heat transport scaling of $Nu \sim O(Ra^{1/2})$, matching the analytical upper-bound scaling and corresponding to the ultimate heat transport regime~\citep{Kraichnan1962,Spiegel1963}. Nevertheless, the physical mechanism that selects these hierarchical structures---and connects them to the underlying stability properties of the flow---remains unclear.

The present work proposes marginal energy stability as the selection principle for the ultimate scaling of $Nu\sim O(Ra^{1/2})$. Here, marginality is not the ordinary marginal linear stability of \cite{Malkus1954} and \cite{Howard1963}, which was later explored by \cite{OConnor2021} and \cite{Wen2022RSTA} for Rayleigh--B\'enard convection and by \cite{Ding2023} for penetrative convection. It is marginal \emph{energy} stability: the saturation of an energy inequality controlling the growth of finite-amplitude perturbations. This distinction is essential. Marginal linear stability concerns neutral eigenmodes of a spectral problem, a framework that provides no transition criterion in shear flows such as pipe flow and plane Couette flow, where the laminar state is linearly stable for all Reynolds numbers; whereas marginal energy stability concerns the point at which the perturbation-energy balance is exactly neutral for the most dangerous admissible disturbance. Here we show that the temperature profiles associated with near-extremal heat flux can be obtained by imposing marginal energy stability. The resulting predictions closely match the best available optimal transport calculations of \cite{Motoki2018} and approach the current best rigorous upper bound of \cite{Plasting2003}. We note, however, that whether the $Ra^{1/2}$ scaling is asymptotically attainable in buoyancy-driven convection remains an open question; the present framework provides constraints consistent with this scaling but does not establish it as the true asymptotic limit.

Importantly, the marginally energy-stable profiles selected by the theory not only predict the largest heat flux compatible with energy stability, but can also be made exact conductive solutions by applying a prescribed internal thermal forcing. Once this forcing is imposed, the same energy-stability condition suppresses convective perturbations. This is potentially important for industrial applications wherein convection is a detrimental factor. For instance, in the semiconductor industry, thermal convection transports oxygen---produced by the dissolution of the quartz crucible---to the growth interface of silicon monocrystals, leading to non-uniform oxygen distribution in the crystal and resulting in inhomogeneous resistivity in the monocrystals \citep{Vegad2014}. Conventional suppression strategies, such as microgravity environments or strong magnetic fields, are either prohibitively expensive or incapable of quenching convection entirely \citep{Potticary2021}. The control mechanism proposed here offers an alternative paradigm: rather than mechanically opposing the flow, it reshapes the thermal background to remove the energy source of convective perturbations.

The remainder of the paper is organized as follows. \S~\ref{sec:formulation} outlines the governing equations, derives the exact perturbation-energy identity and the prescribed-profile minimum-$Ra$ construction, and compares the present framework with the Doering--Constantin and wall-to-wall approaches. \S~\ref{sec:results} presents the numerical results, including the transport law, the marginal-mode hierarchy, and the mean-temperature structure. \S~\ref{sec:control} demonstrates that the marginally energy-stable profile can be enforced dynamically by a prescribed thermal forcing, driving the system to a quiescent conductive state for Rayleigh numbers at or below the design value, while residual convection persists for larger Rayleigh numbers. \S~\ref{sec:conclusions} gives the conclusions of this work.

\section{Governing equations and direct minimum-$Ra$ formulation}\label{sec:formulation}

We consider Boussinesq convection in a horizontally periodic layer $0 \leq z \leq 1$. Using the layer height, the free-fall velocity, and the imposed temperature difference as reference scales, the governing equations are
\begin{subequations}
\label{eq:bouss}
\begin{align}
    \nabla\cdot\bm{u} &= 0,
    \label{eq:gov-cont}\\
    \partial_t\bm{u}+\bm{u}\cdot\nabla\bm{u}
        &= -\nabla p
          +\sqrt{\frac{Pr}{Ra}}\,\nabla^2\bm{u}
          +T\bm{e}_z,
    \label{eq:gov-mom}\\
    \partial_t T+\bm{u}\cdot\nabla T
        &= \frac{1}{\sqrt{Pr\,Ra}}\,\nabla^2 T,
    \label{eq:gov-temp}
\end{align}
\end{subequations}
where $Pr$ is the Prandtl number (the ratio of momentum diffusivity to thermal diffusivity) and $Ra$ is the Rayleigh number (measuring the strength of buoyancy-driven forcing relative to viscous and thermal dissipation). The velocity satisfies no-slip conditions $\bm{u} = \bm{0}$ at $z = 0$ and $z = 1$, the temperature satisfies $T = 1$ at $z = 0$ and $T = 0$ at $z = 1$, and all fields are periodic in the horizontal directions.

We decompose the temperature and velocity as
\begin{equation}
T(\bm{x},t)=\tau(z)+\theta(\bm{x},t),\qquad \bm{u}(\bm{x},t)=\bm{0}+\uu(\bm{x},t),
\label{eq:reynolds-decomposition}
\end{equation}
where $\tau$ is the horizontal and long-time mean temperature, the mean velocity vanishes by horizontal translational symmetry, and $\theta$ and $\uu=(u',v',w')$ are the fluctuations with zero horizontal and temporal mean. The fluctuations satisfy $\theta=0$ and $\uu=\bm{0}$ at the upper and lower boundaries. We define the horizontal and long-time average by
\begin{equation}
    \havg{f}(z) = \lim_{t_*\to\infty}\frac{1}{4t_*L_xL_y}
    \int_0^{t_*}\!\int_{-L_y}^{L_y}\!\int_{-L_x}^{L_x}
    f(x,y,z,t)\,\dd x\,\dd y\,\dd t,
    \label{eq:horizontal-time-average}
\end{equation}
and the full space-time average by
\begin{equation}
    \avg{f} = \int_0^1\havg{f}(z)\,\dd z.
    \label{eq:full-average}
\end{equation}
Averaging equation~\eqref{eq:gov-temp} horizontally and over long times yields the mean temperature equation; subtracting the averaged equations from equations~\eqref{eq:bouss} gives the fluctuation equations:
\begin{subequations}
\label{eq:Reynolds}
\begin{align}
\frac{\dd^2\tau}{\dd z^2} - \sqrt{Pr\,Ra}\,\frac{\dd\havg{w'\theta}}{\dd z} &= 0, \label{eq:meanT-unscaled}\\
    \nabla\cdot\uu &= 0,
    \label{eq:fluct-cont}\\
    \partial_t\uu + \uu\cdot\nabla\uu + \nabla p'
        - \sqrt{\frac{Pr}{Ra}}\,\nabla^2\uu
        - \theta\bm{e}_z
        - \havg{\uu\cdot\nabla\uu}
        &= \bm{0},
    \label{eq:fluct-momentum}\\
    \partial_t\theta + \uu\cdot\nabla\theta
        + w'\frac{\dd\tau}{\dd z}
        - \frac{1}{\sqrt{Pr\,Ra}}\,\nabla^2\theta
        - \frac{\dd\havg{w'\theta}}{\dd z}
        &= 0.
    \label{eq:fluct-temperature}
\end{align}
\end{subequations}

\subsection{Exact perturbation-energy identity}\label{sec:energy-identity}

Taking the scalar product of equation~\eqref{eq:fluct-momentum} with $\uu$, multiplying equation~\eqref{eq:fluct-temperature} by $\theta$, and taking the full space-time average gives
\begin{subequations}
\label{eq:kinetic_energy_balances}
\begin{align}
    \left\langle\uu\cdot\left(
        \partial_t\uu + \uu\cdot\nabla\uu + \nabla p'
        - \sqrt{\frac{Pr}{Ra}}\,\nabla^2\uu
        - \theta\bm{e}_z
        - \havg{\uu\cdot\nabla\uu}
    \right)\right\rangle &= 0,
    \label{eq:kinetic_balances}\\
    \left\langle\theta\left(
        \partial_t\theta + \uu\cdot\nabla\theta
        + w'\frac{\dd\tau}{\dd z}
        - \frac{1}{\sqrt{Pr\,Ra}}\,\nabla^2\theta
        - \frac{\dd\havg{w'\theta}}{\dd z}
    \right)\right\rangle &= 0.
    \label{eq:thermal_balances}
\end{align}
\end{subequations}
The time derivatives vanish for statistically stationary bounded states; the nonlinear transport terms vanish by incompressibility, horizontal periodicity, and the homogeneous wall conditions; the pressure term vanishes after integration by parts; and the subtracted mean terms vanish because $\havg{\uu}=\bm{0}$ and $\havg{\theta}=0$. Using $\avg{\uu\cdot\nabla^2\uu}=-\avg{|\nabla\uu|^2}$ and $\avg{\theta\nabla^2\theta}=-\avg{|\nabla\theta|^2}$, equations~\eqref{eq:kinetic_energy_balances} reduce to
\begin{subequations}
\label{eq:energy-balances}
\begin{align}
    \sqrt{\frac{Pr}{Ra}}\avg{|\nabla\uu|^2} - \avg{w'\theta}
        &= 0,
    \label{eq:kinetic-balance}\\
    \frac{1}{\sqrt{Pr\,Ra}}\avg{|\nabla\theta|^2}
        + \avg{\frac{\dd\tau}{\dd z}\,w'\theta}
        &= 0.
    \label{eq:thermal-balance}
\end{align}
\end{subequations}
Multiplying equation~\eqref{eq:kinetic-balance} by $a$ and adding equation~\eqref{eq:thermal-balance} gives, for every $a>0$,
\begin{equation}
    \avg{
        a\sqrt{\frac{Pr}{Ra}}|\nabla\uu|^2
        + \frac{1}{\sqrt{Pr\,Ra}}|\nabla\theta|^2
        - \left(a-\frac{\dd\tau}{\dd z}\right)w'\theta
    } = 0,
    \label{eq:energy-identity-unscaled}
\end{equation}
where the balance parameter $a$ weights the kinetic part of the quadratic perturbation energy and is not prescribed in the minimum-$Ra$ calculation below.
 
Introducing the Prandtl-rescaled velocity
\begin{equation}
    \ue = \sqrt{Pr}\,\uu, \qquad \we = \sqrt{Pr}\,w',
    \label{eq:pr-rescaling}
\end{equation}
removes the explicit dependence on $Pr$.  Equations~\eqref{eq:meanT-unscaled} and~\eqref{eq:energy-identity-unscaled} become
\begin{equation}
    \frac{\dd^2\tau}{\dd z^2}
    - \sqrt{Ra}\,\frac{\dd\havg{\we\theta}}{\dd z} = 0
    \label{eq:tau}
\end{equation}
and
\begin{equation}
    \avg{
        \frac{a}{\sqrt{Ra}}|\nabla\ue|^2
        + \frac{1}{\sqrt{Ra}}|\nabla\theta|^2
        - \left(a-\frac{\dd\tau}{\dd z}\right)\we\theta
    } = 0.
    \label{eq:energy_identity}
\end{equation}
Equations~\eqref{eq:tau} and~\eqref{eq:energy_identity} are exact: they follow directly from the governing equations without approximation. Equation~\eqref{eq:energy_identity} serves as the starting point for the minimum-Rayleigh-number problem for a prescribed mean temperature field $\tau(z)$.

\subsection{Rayleigh quotient and the neutral state}
\label{sec:minRa}

Define the dissipation and production forms
\begin{subequations}
\label{eq:AB_definitions}
\begin{align}
    \mathcal{A}_a[X]
        &= \avg{a|\nabla\ue|^2+|\nabla\theta|^2},
    \label{eq:A_def}\\
    \mathcal{B}_{\tau,a}[X]
        &= \avg{\left(a-\frac{\dd\tau}{\dd z}\right)\we\theta},
    \label{eq:B_def}
\end{align}
\end{subequations}
where $X=(\ue,\theta)$, $\nabla\cdot\ue=0$, and $\ue=\bm{0}$, $\theta=0$ at the upper and lower boundaries. For any neutral disturbance with $\mathcal{B}_{\tau,a}>0$, equation~\eqref{eq:energy_identity} gives
\begin{equation}
    \sqrt{Ra} = \frac{\mathcal{A}_a[X]}{\mathcal{B}_{\tau,a}[X]}.
    \label{eq:Ra_quotient}
\end{equation}
The mean temperature profile $\tau(z)$ is now prescribed; the balance parameter $a$ is an unknown positive scalar determined together with the neutral disturbance. For a given prescribed $\tau$, the Nusselt number is consequently given, and there is a minimal $Ra$ that delivers the largest heat-flux scaling exponent $\alpha$ in the power-law ansatz $Nu\sim Ra^\alpha$ (see also the piecewise analysis in Appendix~\ref{app:piecewise}). We therefore define
\begin{equation}
    \sqrt{Ra_{\min}[\tau]}
    = \min_{\substack{a>0,\;X\ne\bm{0},\;\nabla\cdot\ue=0\\
    \mathcal{B}_{\tau,a}[X]>0}}
    \frac{\mathcal{A}_a[X]}{\mathcal{B}_{\tau,a}[X]}.
    \label{eq:Ra_min_main}
\end{equation}
Equation~\eqref{eq:Ra_min_main} is the smallest Rayleigh number compatible with the exact perturbation-energy balance for the prescribed mean field.

For later use, define
\begin{equation}
    \mathcal{Q}_{Ra}[\tau,a;X]
    := \frac{1}{\sqrt{Ra}}\mathcal{A}_a[X]-\mathcal{B}_{\tau,a}[X].
    \label{eq:energy_constraint_main}
\end{equation}
The joint minimum~\eqref{eq:Ra_min_main} admits the equivalent characterization
\begin{equation}
    Ra \le Ra_{\min}[\tau]
    \quad\Longleftrightarrow\quad
    \mathcal{Q}_{Ra}[\tau,a;X] \ge 0
    \quad\text{for every }a>0\text{ and every admissible }X.
    \label{eq:minRa_spectral_equivalence}
\end{equation}
At marginality, equality is attained by at least one pair $(a_*,X_*)$, with $a$ selected by the active minimum rather than specified in advance.
When $Ra<Ra_{\min}$, $\mathcal{Q}_{Ra}>0$ and the system is energy stable: all perturbations about the conductive state $T=\tau(z)$ decay and no convective motion can be sustained. Conversely, any flow governed by the Boussinesq equations with $\uu\ne\bm{0}$ satisfies $Ra>Ra_{\min}$. The present study therefore focuses on the marginal state at the edge of energy stability. The marginal modes satisfy the stationarity conditions of the Lagrangian functional
\begin{equation}
    \mathcal{L} := \mathcal{Q}_{Ra} - \avg{P(\bm{x})\,\nabla\cdot\ue},
    \label{eq:minRa_Lagrange}
\end{equation}
where $P(\bm{x})$ is a Lagrange-multiplier field enforcing incompressibility.

Stationarity of $\mathcal{L}$ with respect to the disturbance fields and the balance parameter $a$ gives
\begin{subequations}
\label{eq:el}
\begin{align}
    \nabla\cdot\ue &= 0,
    \label{eq:incomp_el}\\
    -\frac{2a}{\sqrt{Ra}}\nabla^2\ue
        - \left(a-\frac{\dd\tau}{\dd z}\right)\theta\bm{e}_z + \nabla P
        &= \bm{0},
    \label{eq:momentum_el}\\
    -\frac{2}{\sqrt{Ra}}\nabla^2\theta
        - \left(a-\frac{\dd\tau}{\dd z}\right)\we
        &= 0,
    \label{eq:theta_el}\\
    \avg{|\nabla\ue|^2} - \sqrt{Ra}\avg{\we\theta}
        &= 0.
    \label{eq:a_el}
\end{align}
\end{subequations}
Note that equations~\eqref{eq:el} determine $a$, the marginal wavenumbers, and the shapes of the marginal disturbances, but not their physical amplitudes.

\subsection{The mean temperature profile at marginal energy stability}
\label{sec:profile-selection}

The stationarity conditions~\eqref{eq:el} enforce marginal energy stability on the mean temperature profile, but leave the amplitudes of the marginal disturbances undetermined. To close the system, we additionally impose the exact mean heat equation~\eqref{eq:tau}: the modal amplitudes are fixed by requiring that their aggregate heat flux is consistent with the profile through equation~\eqref{eq:tau}. These two requirements---marginal energy stability and satisfaction of the mean heat equation---jointly select a unique mean temperature profile. Equations~\eqref{eq:tau} and~\eqref{eq:el}, together with the boundary conditions, form the closed profile-mode system.

Appendix~\ref{app:background-lagrangian} shows that equations~\eqref{eq:tau} and~\eqref{eq:el} are the stationarity conditions of an auxiliary convex-concave background Lagrangian, establishing uniqueness of the optimal mean profile. When a solution satisfying $\mathcal{Q}_{Ra}\ge0$ is obtained, it is globally optimal and the Nusselt number is
\begin{equation}
    Nu = -\left.\frac{\dd\tau}{\dd z}\right|_{z=0}.
    \label{eq:nu}
\end{equation}

We now compare the present formulation with the Doering--Constantin background method and wall-to-wall optimal transport.
\begin{enumerate}
\item \textit{Doering--Constantin background method.}
At prescribed $Ra$, the background method varies an auxiliary temperature profile and associated balance parameters to obtain a rigorous upper bound on $Nu$ \citep{DC1996,DingKerswell20}. The result is a universal inequality valid for all admissible solutions. Our marginal-energy-stability constraint $\mathcal{Q}_{Ra}\ge0$ resembles the spectral constraint in upper-bound theory, albeit in different forms. The key difference is that $\tau$ in the present work denotes the mean temperature and satisfies the exact mean heat equation, whereas the background temperature in upper-bound theory is an auxiliary function that need not satisfy any physical equation.
 
\item \textit{Wall-to-wall optimal transport.}
Wall-to-wall theory prescribes a measure of flow intensity, such as kinetic energy or enstrophy, and optimizes over divergence-free velocity fields to maximize heat transfer \citep{Doering2019}. The scalar field satisfies the full advection-diffusion equation, and the mean heat equation is therefore naturally satisfied.  Moreover, equation~\eqref{eq:a_el} is equivalent to the wall-to-wall formulation when the enstrophy is fixed. We expect our solution to share many features with wall-to-wall theory, particularly the three-dimensional results of \cite{Motoki2018}.
\end{enumerate}

\subsection{Numerical method}\label{sec:numerical-method}

The system is solved by a Fourier--Chebyshev spectral method \citep{Ding2024}. Newton--Raphson is applied to solve the resulting nonlinear algebraic system, and pseudo-arclength continuation follows the branch connected to conductive onset.
 
By a Squire-type reduction, the marginal modes can be represented by two-dimensional disturbances. The numerical strategy uses the generalized energy-stability eigenproblem and follows related spectral-constraint calculations \citep{Plasting2003,DingKerswell20,Ding2024}. We therefore expand the fields in horizontal Fourier modes as
\begin{equation}
    \tilde{u} = \sum_{i=1}^{\infty}\hat{u}_i(z)\cos(k_i x) \quad \text{and}
    \quad
    (\we,\theta,P) = \sum_{i=1}^{\infty}
    \bigl(\hat{w}_i(z),\hat{\theta}_i(z),\hat{P}_i(z)\bigr)\sin(k_i x).
    \label{eq:S_fourier_expansion}
\end{equation}
Substitution into equations~\eqref{eq:tau} and~\eqref{eq:el} gives
\begin{subequations}
\label{eq:S_mode}
\begin{align}
    \frac{\dd^2\tau}{\dd z^2}
        - \frac{\sqrt{Ra}}{2}\sum_{i=1}^{\infty}
          \frac{\dd(\hat{w}_i\hat{\theta}_i)}{\dd z}
        &= 0,
    \label{eq:S_mode_meanT}\\
    -k_i\hat{u}_i+\frac{\dd\hat{w}_i}{\dd z}
        &= 0,
    \label{eq:S_mode_continuity}\\
    -\frac{2a}{\sqrt{Ra}}\left(\frac{\dd^2\hat{u}_i}{\dd z^2}
        -k_i^2\hat{u}_i\right)+k_i\hat{P}_i
        &= 0,
    \label{eq:S_mode_xmom}\\
    -\frac{2a}{\sqrt{Ra}}\left(\frac{\dd^2\hat{w}_i}{\dd z^2}
        -k_i^2\hat{w}_i\right)
        -\left(a-\frac{\dd\tau}{\dd z}\right)\hat{\theta}_i
        +\frac{\dd\hat{P}_i}{\dd z}
        &= 0,
    \label{eq:S_mode_zmom}\\
    -\frac{2}{\sqrt{Ra}}\left(\frac{\dd^2\hat{\theta}_i}{\dd z^2}
        -k_i^2\hat{\theta}_i\right)
        -\left(a-\frac{\dd\tau}{\dd z}\right)\hat{w}_i
        &= 0,
    \label{eq:S_mode_heat}\\
    \sum_{i=1}^{\infty}\left[
        \int_0^1\!\left\{
            \left(\frac{\dd\hat{u}_i}{\dd z}\right)^2
            +k_i^2\hat{u}_i^2
            +\left(\frac{\dd\hat{w}_i}{\dd z}\right)^2
            +k_i^2\hat{w}_i^2
        \right\}\dd z
        -\sqrt{Ra}\int_0^1\hat{w}_i\hat{\theta}_i\,\dd z
    \right] &= 0.
    \label{eq:S_mode_marginality}
\end{align}
\end{subequations}
System~\eqref{eq:S_mode} admits non-trivial solutions for disturbance fields only at a discrete set of critical wavenumbers $k_{c,i}$ ($i=1,\dots,M$), determined as part of the solution.
After substituting the Fourier expansion~\eqref{eq:S_fourier_expansion}, $\mathcal{L}$ depends explicitly on $k_i$, and stationarity with respect to $k_i$ gives the additional algebraic equation
\begin{equation}
\frac{2k_i}{\sqrt{Ra}}\int_0^1
      \left[a\left(\hat{u}_i^2+\hat{w}_i^2\right)+\hat{\theta}_i^2\right]\dd z
    + \int_0^1\hat{u}_i\hat{P}_i\,\dd z = 0.
    \label{eq:S_wavenumber_condition}
\end{equation}

The differential-integral system is discretized using Chebyshev polynomials,
\begin{equation}
    (\hat{u}_i,\hat{w}_i,\hat{\theta}_i,\hat{P}_i)(z)
    = \sum_{n=0}^{N}
      (\tilde{u}_{in},\tilde{w}_{in},\tilde{\theta}_{in},\tilde{P}_{in})
      T_n(2z-1),
\end{equation}
where $T_n(\xi)=\cos(n\cos^{-1}\xi)$ and $\xi=2z-1$. Solutions are continued from the bifurcation point of the conduction state $\tau=1-z$ at $Ra_c\simeq1708$.

New marginal modes are detected by solving the corresponding eigenvalue problem
\begin{subequations}
\label{eq:eigen}
\begin{align}
    -k_i\hat{u}_i+\frac{\dd\hat{w}_i}{\dd z}
        &= 0,
    \label{eq:eigen1}\\
    -\frac{2a}{\sqrt{Ra}}\left(\frac{\dd^2\hat{u}_i}{\dd z^2}
        -k_i^2\hat{u}_i\right)+k_i\hat{P}_i
        &= \lambda\hat{u}_i,
    \label{eq:eigen2}\\
    -\frac{2a}{\sqrt{Ra}}\left(\frac{\dd^2\hat{w}_i}{\dd z^2}
        -k_i^2\hat{w}_i\right)
        -\left(a-\frac{\dd\tau}{\dd z}\right)\hat{\theta}_i
        +\frac{\dd\hat{P}_i}{\dd z}
        &= \lambda\hat{w}_i,
    \label{eq:eigen3}\\
    -\frac{2}{\sqrt{Ra}}\left(\frac{\dd^2\hat{\theta}_i}{\dd z^2}
        -k_i^2\hat{\theta}_i\right)
        -\left(a-\frac{\dd\tau}{\dd z}\right)\hat{w}_i
        &= \lambda\hat{\theta}_i.
    \label{eq:eigen4}
\end{align}
\end{subequations}
Since the marginal-energy-stability constraint $\mathcal{Q}_{Ra}\ge0$ is equivalent to $\lambda\ge0$, when a negative eigenvalue is found, the corresponding mode is added to equation~\eqref{eq:S_mode_meanT}; the mean temperature profile then readjusts until all eigenvalues are non-negative.

\begin{figure}
    \centering
    \safeincludegraphics[width=0.72\linewidth]{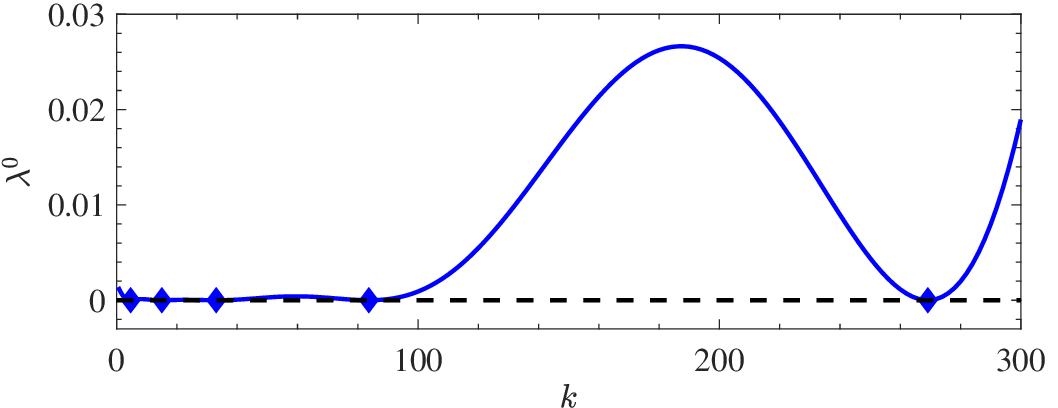}
    \caption{Ground-state eigenvalue $\lambda^0$ of equations~\eqref{eq:eigen} at $Ra=10^8$. The spectrum is non-negative everywhere and touches zero (dashed line) exclusively at five critical wavenumbers $k_{c,1}$--$k_{c,5}$ (diamonds); the number of critical wavenumbers $k_c$ depends on $Ra$ (see figure~\ref{fig:transport}$b$).}
    \label{fig:eigen1e8}
\end{figure}

Two independent checks verify the accuracy of the computed solution. First, resolution convergence is assessed using the nonlinear profile-mode system~\eqref{eq:S_mode}. At $Ra=10^8$, the computed Nusselt number~\eqref{eq:nu} changes from $245.3003$ with $N=200$ Chebyshev modes to $245.3023$ with both $N=250$ and $N=300$, corresponding to a relative difference of approximately $8.2\times10^{-6}$ (table~\ref{tab:S_convergence}). For $Ra\le10^8$, $N=250$ Chebyshev modes are sufficient to resolve the boundary layers.  Second, the eigenvalue problem~\eqref{eq:eigen} is scanned over a broad wavenumber interval using the converged profile $\tau(z)$; the resulting spectrum remains non-negative throughout and vanishes only at the critical wavenumbers (figure~\ref{fig:eigen1e8}). These checks confirm, respectively, convergence of the nonlinear solution and satisfaction of the marginal-energy-stability constraint. 

\begin{table}
    \centering
    \caption{Convergence of the Nusselt number with Chebyshev resolution
        at $Ra=10^8$.}
    \label{tab:S_convergence}
    \begin{tabular}{lccc}
        \toprule
        Chebyshev modes $N$ & 200    & 250      & 300      \\
        \midrule
        $Nu$                & 245.3003 & 245.3023 & 245.3023 \\
        \bottomrule
    \end{tabular}
\end{table}

\section{Heat transport limited by marginal energy stability}
\label{sec:results}

\subsection{Heat transport and marginal modes}

\begin{figure}
  \centering
  \safeincludegraphics[width=0.7\linewidth]{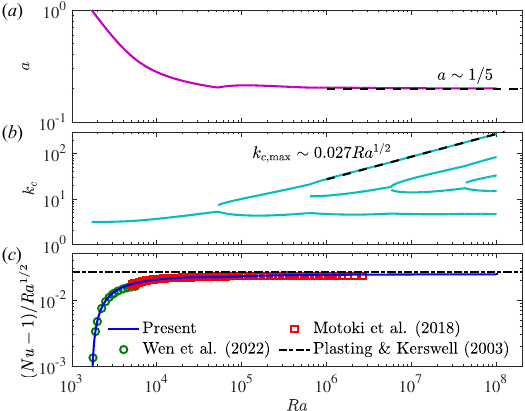}
	\caption{Variation of the balance parameter $a$, the critical wavenumbers $k_c$, and the compensated heat transport $Nu$ as functions of the Rayleigh number $Ra$.  ($a$) Dependence of $a$ on $Ra$. The dashed line indicates the asymptotic limit $a\sim1/5$ as $Ra\to\infty$.  ($b$)  Bifurcation diagram of $k_c$, illustrating the emergence of additional marginal modes as $Ra$ increases.  The largest critical wavenumber follows $k_{c,\max} \sim 0.027Ra^{1/2}$ (dashed line).  ($c$)  Comparison of heat transport predictions. Green circles denote optimal exact steady solutions of the full Boussinesq equations computed by \citet{Wen2022JFM}.  Red squares denote the maximal transport obtained by \citet{Motoki2018} from the wall-to-wall optimal transport problem under a fixed-enstrophy constraint; the original data, given as $Nu$ versus $Pe= \sqrt{Ra(Nu-1)}$, have been converted to $Nu$ versus $Ra$.  The dash-dotted line indicates the rigorous upper bound $(Nu-1)\le 0.02634Ra^{1/2}$ reported by \citet{Plasting2003}.
	}\label{fig:transport}
\end{figure}

The marginal energy-stability formulation reproduces the classical onset of convection. For $Ra < Ra_c \simeq 1708$, the conductive state ($T=1-z$, $\bm{u}=\bm{0}$) is globally energy stable. At $Ra=Ra_c$, the first energy-neutral mode emerges, and the profile selected by the self-consistent minimum-$Ra$ procedure departs from the linear conductive profile.  As $Ra$ increases, the balance parameter $a$ decreases from unity near onset and asymptotically approaches $1/5$ in the large-$Ra$ limit (figure~\ref{fig:transport}$a$).

The energy-neutral modes possess a discrete spectral structure. For each $Ra$, only certain horizontal wavenumbers saturate the energy-stability constraint with respect to the selected profile $\tau(z)$. With increasing $Ra$, additional energy-neutral modes appear, corresponding to progressively finer structures near the thermal boundary layers. The largest critical wavenumber follows the scaling
\begin{equation}
 k_{c,\max}\sim0.027Ra^{1/2},
\label{eq:kc_scaling}
\end{equation}
as shown in figure~\ref{fig:transport}($b$). This scaling reflects the development of increasingly thin energy-stability-limited boundary-layer structures.

The heat transport predicted by the marginal energy-stability framework agrees closely with independent optimal-transport results (figure~\ref{fig:transport}$c$). At moderate $Ra$, the predictions nearly coincide with the optimal steady solutions of the full Boussinesq equations computed by \citet{Wen2022JFM}.  At larger $Ra$, they remain close to the wall-to-wall optimal-transport values obtained under a fixed-enstrophy constraint by \citet{Motoki2018}.  In the asymptotic regime, the marginal energy-stability theory predicts
\begin{equation}
Nu\sim0.0245Ra^{1/2},
\label{eq:asymptotic_nu}
\end{equation}
which lies within a few percent of the wall-to-wall scaling $Nu\sim0.0235Ra^{1/2}$ \citep{Motoki2018} and below the rigorous upper bound $(Nu-1)\leq0.02634Ra^{1/2}$ \citep{Plasting2003}.  The proximity of these independent calculations suggests that the energy-stability constraint captures an important part of the mechanism limiting the heat flux.

The piecewise-profile calculation in Appendix~\ref{app:piecewise} provides an analytical interpretation of these numerical scalings.  It yields
\begin{equation}
\delta\gtrsim \frac{16\sqrt{a}}{1+a}Ra^{-1/2},
\qquad
Nu\leq C_{\mathrm{pw}}(a)\,Ra^{1/2},
\qquad
C_{\mathrm{pw}}(a)=\frac{(1+a)^2}{32\sqrt{a}},
\label{eq:piecewise_main_comparison}
\end{equation}
where $\delta$ is the conductive-layer thickness and $C_{\mathrm{pw}}$ is the coefficient of the $Nu$--$Ra$ scaling.  The estimate predicts $\delta=O(Ra^{-1/2})$, consistent with the collapse of the numerical profiles in the coordinate $zRa^{1/2}$ and with the observed $k_{c,\max}=O(Ra^{1/2})$.  Because $a$ is included in the prescribed-profile minimization, its asymptotic value is selected together with the marginal modes. The coupled profile-mode branch approaches $a=1/5$ in figure~\ref{fig:transport}($a$), consistent with the analysis in Appendix~\ref{app:piecewise}.  Evaluating the piecewise coefficient at this numerical limit gives $C_{\mathrm{pw}}=9\sqrt{5}/200\simeq0.1006$, about $4.1$ times the numerical coefficient $0.0245$.  Thus the piecewise calculation explains the exponent and boundary-layer scale, while the smooth multi-mode closure determines the balance parameter and the quantitative transport coefficient.

\begin{figure}
\centering
\safeincludegraphics[width=0.7\linewidth]{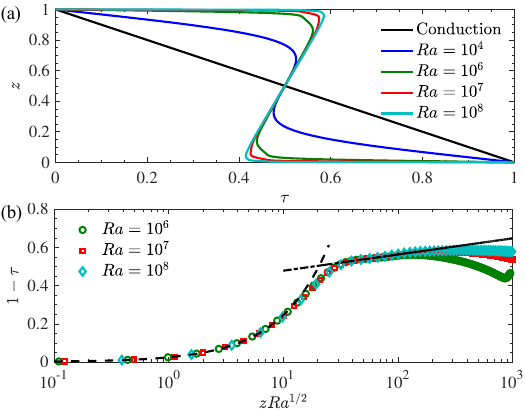}
	\caption{Mean temperature profiles predicted by the marginal energy-stability theory and their near-wall scaling. $(a)$ Dependence of $\tau(z)$ on $Ra$. $(b)$ Scaled spatial structure of $1-\tau$ near the lower boundary. At large $Ra$, the mean temperature profile exhibits a clear multi-layer structure. Near the lower boundary, conduction dominates within an inner boundary layer characterized by $zRa^{1/2}<10$. Away from the wall, a logarithmic-like layer emerges for $30\lesssim zRa^{1/2}\lesssim200$. In the bulk region, $\tau$ displays a stably stratified structure. Both the conductive inner layer and the logarithmic-like layer exhibit self-similar behaviour and are fitted by $1-\tau\approx zNu\approx0.0245zRa^{1/2}$ (dashed line) and $1-\tau\approx0.0368\ln(zRa^{1/2})+0.393$ (dash-dot line), respectively. By symmetry, the same structures appear near the upper boundary.
	}\label{fig:profiles}
\end{figure}

\subsection{Mean-temperature structure}

The optimal mean profiles reveal how energy-stability saturation organizes extreme transport. Figure~\ref{fig:profiles} shows $\tau(z)$ for increasing $Ra$. Above onset, the profile develops sharp thermal boundary layers near the walls and a stably stratified interior. At large $Ra$, the lower boundary region separates into two distinct layers. Very close to the wall, conduction dominates and the profile remains nearly linear,
\begin{equation}
1-\tau\simeq zNu\simeq0.0245zRa^{1/2},
\label{eq:inner_layer}
\end{equation}
for $zRa^{1/2}\lesssim10$.  Farther from the wall, an intermediate logarithmic-like layer emerges,
\begin{equation}
1-\tau\simeq0.0368\ln(zRa^{1/2})+0.393,
\label{eq:log_layer}
\end{equation}
for $30\lesssim zRa^{1/2}\lesssim200$, which is very close to the profile computed from three-dimensional wall-to-wall theory by \citet{Motoki2018}. By symmetry, analogous boundary structures occur near the upper wall ($z=1$). 

The computed profile contains a conductive inner layer, a logarithmic-like intermediate layer, and a stably stratified core. The piecewise profile of Appendix~\ref{app:piecewise} is the simplest asymptotic representation of the first and third of these regions: its steep wall segments represent the $O(Ra^{-1/2})$ conductive layers, while its positive interior slope $a$ represents the weak stable stratification (the interior slope of $\tau$ is $1/5$, consistent with the asymptotic value $a\sim1/5$). The piecewise profile does not resolve the logarithmic-like transition, which is generated by the hierarchy of marginal modes in the numerical solution. These features closely resemble the structural layers observed in wall-to-wall optimal transport solutions \citep{Motoki2018}. Crucially, this agreement extends beyond the global heat flux to the detailed spatial organization of the mean profiles, implying that the energy-stability and wall-to-wall formulations select related thermal balances. The close agreement of the present heat flux and mean-temperature structure with the three-dimensional wall-to-wall results \citep{Motoki2018} may therefore indicate a common physical transport mechanism rather than mathematical coincidence.

\section{Convection suppression by prescribed thermal forcing}\label{sec:control}

The marginally energy-stable profiles are selected by the minimum-$Ra$ marginal-mode closure, but they can also be realized dynamically. Consider modifying the temperature equation~\eqref{eq:gov-temp} by introducing a prescribed internal thermal forcing on the right-hand side,
\begin{equation}
    s(z) = -\frac{1}{\sqrt{Pr\,Ra}}\frac{\dd^2 T_m}{\dd z^2},
    \label{eq:forcing}
\end{equation}
where $T_m(z)$ is the marginally energy-stable mean temperature profile obtained from equations~\eqref{eq:S_mode} at the given $Ra$. Under this forcing, the background profile $T=T_m(z)$ and motionless velocity field $\bm{u}=\bm{0}$ constitute an exact steady conductive solution of the modified Boussinesq system, while the perturbation energy identity~\eqref{eq:energy_identity} for disturbances about $T_m$ retains the same form as in the unforced system, since the forcing enters only through the mean temperature and not the perturbation equations. Because $T_m$ satisfies the marginal energy-stability condition, perturbations cannot extract net energy from the background state, and the system relaxes to the motionless conductive state $T=T_m(z)$, $\bm{u}=\bm{0}$.

The effectiveness of the control depends on the Rayleigh number $Ra$ relative to the design value $Ra_c$ at which $T_m$ was computed. Since $T_m$ is constructed so that $Ra_{\min}[T_m]=Ra_c$, the characterization~\eqref{eq:minRa_spectral_equivalence} gives two distinct cases:
\begin{enumerate}
    \item For $Ra\le Ra_c$: $\mathcal{Q}_{Ra}\ge0$ for all admissible disturbances. The background state $T=T_m(z)$, $\bm{u}=\bm{0}$ is globally energy stable; all perturbations decay and the flow converges to the quiescent conductive state.
    \item For $Ra>Ra_c$: there exist disturbances for which $\mathcal{Q}_{Ra}<0$. The energy-stability condition is violated; convective motion may be sustained and the forcing may not be able to fully suppress the flow.
\end{enumerate}
This control mechanism differs fundamentally from direct velocity suppression. Rather than opposing the flow field pointwise, the forcing reshapes the thermal background to remove the energy source of convective perturbations, driving the system toward the conductive state $T=T_m(z)$, $\bm{u}=\bm{0}$.

\begin{figure}
    \centering
    \safeincludegraphics[width=0.7\linewidth]{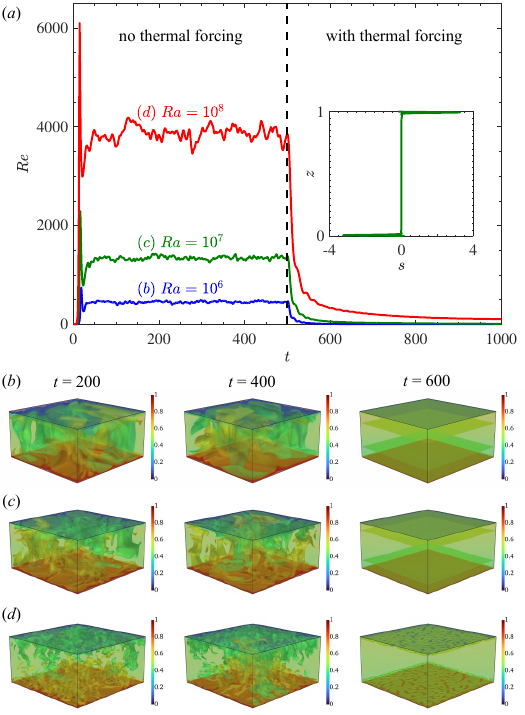}
    \caption{Suppression of convection by prescribed thermal forcing. $(a)$ Instantaneous Reynolds number $Re$~\eqref{eq:reynolds} from three-dimensional DNS at $Pr=1$ for $Ra=10^6$ (blue), $Ra=10^7$ (green), and $Ra=10^8$ (red). The thermal forcing $s(z)$~\eqref{eq:forcing} is activated at $t=500$ (vertical dashed line); the inset shows the corresponding forcing profile. The marginally energy-stable background profile $T_m(z)$ computed at $Ra_0=10^7$ is used for all three simulations. $(b)$--$(d)$ Representative temperature fields at $t=200$, $400$, and $600$ for $Ra=10^6$, $Ra=10^7$, and $Ra=10^8$, respectively. For $Ra\le Ra_c=10^7$, the flow transitions to the quiescent conductive state; for $Ra=10^8>Ra_c$, residual convective motion persists.}
    \label{fig:control}
\end{figure}

We verify this picture through three-dimensional DNS at $Ra=10^6$, $10^7$, and $10^8$ with $Pr=1$, in the horizontally periodic domain $(x,y,z)\in[0,2]\times[0,2]\times[0,1]$, using a Fourier--Chebyshev spectral solver in Dedalus \citep{Burns2020}. A second-order Runge--Kutta scheme is used for time integration and the $3/2$ dealiasing rule is applied. All three simulations use a resolution of $256^3$. Each case is initialized from the linear conductive state with a small random thermal perturbation and no internal forcing, and develops convective motion before the control is applied. We monitor the instantaneous volume-averaged root-mean-square Reynolds number
\begin{equation}
    Re = \sqrt{\frac{Ra}{Pr}}
    \left(\frac{1}{V}\int_V|\bm{u}|^2\,\dd V\right)^{1/2},
    \label{eq:reynolds}
\end{equation}
as shown in figure~\ref{fig:control}($a$). At $t=500$, the thermal forcing~\eqref{eq:forcing} is activated in all three simulations using the same $T_m$ computed at $Ra_c=10^7$. For $Ra=10^6$ and $Ra=10^7$, both at or below the design value, the Reynolds number decays to zero, confirming that the profile-based control drives the system to the globally energy-stable quiescent conductive state. For $Ra=10^8$, which exceeds $Ra_0$, the Reynolds number decays substantially but does not reach zero, consistent with the theoretical prediction that the energy-stability condition is violated and convective motion can persist. The temperature fields in panels $(b)$--$(d)$ of figure~\ref{fig:control} illustrate the response across all three Rayleigh numbers: the flow becomes quiescent for $Ra\le Ra_c$ and retains residual convective motion for $Ra>Ra_c$.

\section{Discussion and conclusions}\label{sec:conclusions} 
 
Classical upper-bound and optimal-transport theories identify absolute limits on the Nusselt number $Nu$, but their extremal states often lack a direct dynamical interpretation from the viewpoint of stability theory. The present study proposes marginal energy stability as the organizing principle for those extremal states. The mean temperature profile selected by marginal energy stability is the unique one for which buoyant production, viscous dissipation, and thermal diffusion reach an energy-neutral balance, in contrast to the classical $Nu\sim Ra^{1/3}$ scaling derived from Malkus's marginal linear stability assumption.
 
Turbulence intensity is not the fundamental quantity that determines the largest attainable heat flux by convection. A strongly energy-stable profile suppresses motion and cannot support large advective transport, whereas a strongly energy-unstable profile permits perturbation growth but does not represent an energy-controlled extremal balance. The marginally energy-stable profiles identified here lie between these limits: they sustain the largest buoyancy production compatible with the energy-stability constraint. This explains why the predicted heat flux, $Nu\sim0.0245Ra^{1/2}$ at large $Ra$, lies close to the wall-to-wall optimal transport and the strongest available rigorous bounds, and why the corresponding profiles contain the same conductive inner layers, logarithmic-like intermediate layers, and stably stratified bulk observed in wall-to-wall theory.
 
A further consequence of the present framework concerns the asymptotic scaling of heat transport. Within marginal energy stability theory, $Ra_{\min}[\tau]$ is the smallest Rayleigh number compatible with the exact perturbation-energy balance for any prescribed mean profile $\tau$. It follows that any turbulent flow achieving a given $Nu$ must satisfy $Ra>Ra_{\min}[\tau]$. In the power-law ansatz $Nu\sim cRa^{\alpha}$, this implies either $\alpha<1/2$ or $c<0.0245$. Rigorous proof of both cases is challenging: the former would establish that the $Ra^{1/2}$ regime is not asymptotically attained, while the latter implies that $c$ may be a decreasing function of $Ra$. Within the framework of marginal energy stability, we expect to establish $\alpha<1/2$ in two-dimensional Rayleigh--B\'enard convection between two stress-free boundaries by incorporating the vorticity equation as an additional constraint \citep{Whitehead}.
 
A second consequence is that near-optimal transport configurations can be made physically realizable without turbulent mixing. By imposing an internal thermal forcing that maintains the marginally energy-stable profile, the system is driven toward a conductive state with zero velocity. This control mechanism does not mechanically oppose the flow; rather, it reshapes the thermal background to remove the energy source of convective perturbations. Direct numerical simulations confirm this idea: for $Ra$ at or below the design value $Ra_c$, an initially turbulent flow relaxes to the quiescent conductive state once the energy-stability-limited profile is imposed, demonstrating that convective motion can be suppressed by thermal background shaping alone. This control principle offers an alternative paradigm for systems where convective motion is detrimental. In microchannel thermal management, buoyancy can induce temperature non-uniformities and hot spots \citep{Qu2002}. In crystal growth, convection drives compositional inhomogeneities and structural defects \citep{Vegad2014}, while existing mitigation strategies such as magnetic damping remain energy intensive and incomplete \citep{Potticary2021}. Spatially distributed thermal forcing---potentially via selective optical cooling or heating \citep{Shenhav2019Patent}---could approximate the required profile, suppressing convective instabilities. While experimental implementation must address finite-resolution forcing, material constraints, and three-dimensional boundary effects, stabilization is achieved globally through the energy budget rather than through localized mechanical damping, which makes the underlying mechanism robust.
 
More broadly, these findings suggest that energy-stability saturation provides a general route for interpreting variational transport limits in driven dissipative systems. Many nonequilibrium flows are constrained by mathematical bounds whose extremal configurations are difficult to connect to dynamics. This work shows that profiles approaching such absolute limits can emerge from a physically meaningful energy-stability constraint, and that marginal energy stability acts not only as a constraint on admissible dynamics but as a design principle for controlling transport far from equilibrium.

\section*{Acknowledgements}
The authors thank B. Zhang and Z. Ouyang for providing the direct numerical simulation data, G. Kawahara for sharing numerical data from the wall-to-wall optimal-transport approach, and J. J. Tao, B. Hof and C. Sun for helpful discussions.
ZD was supported by the project ``Investigation into Turbulence Transport in Spheres under Multiphysics Fields'' (grant KJZ-YY-NLT0604) and the National Natural Science Foundation of China (grant 52176065).

The authors report no conflict of interest.

\appendix

\section{Piecewise-profile estimate}\label{app:piecewise}

In this appendix, we use a piecewise-linear mean temperature profile to derive an analytical estimate of the transport scaling subject to the marginal energy-stability constraint. The profile is taken to be symmetric,
\begin{equation}
\tau(z)=
\begin{cases}
1-\dfrac{1+a(1-2\delta)}{2\delta}z,
    & 0\le z<\delta,\\[6pt]
a(z-\delta)+\dfrac{1-a(1-2\delta)}{2},
    & \delta\le z<1-\delta,\\[6pt]
-\dfrac{1+a(1-2\delta)}{2\delta}(z-1),
    & 1-\delta\le z\le1.
\end{cases}
\label{eq:S_piecewise_tau}
\end{equation}
For this prescribed profile, equation~\eqref{eq:Ra_min_main} determines the least neutral Rayleigh number by minimizing jointly over the admissible disturbances and the unknown parameter $a>0$. Since $a-\dd\tau/\dd z=0$ in the bulk $\delta\le z\le1-\delta$ and equals $(1+a)/(2\delta)$ in each boundary layer, substitution of the piecewise profile gives
\begin{equation}
\avg{\left(a-\frac{\dd\tau}{\dd z}\right)\we\theta}
=
\frac{1+a}{2\delta}
\left(
\int_0^\delta \havg{\we\theta}\,\dd z
+\int_{1-\delta}^{1}\havg{\we\theta}\,\dd z
\right).
\label{eq:S_piecewise_transport}
\end{equation}

To bound the production term, we estimate each boundary-layer integral. The triangle inequality gives
\begin{equation}
\left|
\avg{\left(a-\frac{\dd\tau}{\dd z}\right)\we\theta}
\right|
\le
\frac{1+a}{2\delta}
\left(
\left|\int_0^\delta\havg{\we\theta}\,\dd z\right|
+
\left|\int_{1-\delta}^{1}\havg{\we\theta}\,\dd z\right|
\right).
\label{eq:S_piecewise_bound1}
\end{equation}
Using the homogeneous boundary conditions of $\we$ and $\theta$, the lower boundary-layer contribution can be written as
\begin{equation}
\left|\int_0^\delta\havg{\we\theta}\,\dd z\right|
=
\left|
\int_0^\delta
\left(\int_0^z\partial_{z'}\we\,\dd z'\right)
\left(\int_0^z\partial_{z'}\theta\,\dd z'\right)\dd z
\right|.
\end{equation}
The Cauchy--Schwarz inequality gives
\begin{equation}
\left|\int_0^z\partial_{z'}\we\,\dd z'\right|
\le
\sqrt{\frac{z}{a}}
\left(\int_0^z a|\partial_{z'}\we|^2\,\dd z'\right)^{1/2}
\end{equation}
and
\begin{equation}
\left|\int_0^z\partial_{z'}\theta\,\dd z'\right|
\le
2\sqrt{z}
\left(\int_0^z\frac{1}{4}|\partial_{z'}\theta|^2\,\dd z'\right)^{1/2}.
\end{equation}
Applying Young's inequality yields
\begin{equation}
\left|\int_0^\delta\havg{\we\theta}\,\dd z\right|
\le
\frac{\delta^2}{2\sqrt{a}}
\left(
\int_0^\delta a|\partial_z\we|^2\,\dd z
+\int_0^\delta\frac{1}{4}|\partial_z\theta|^2\,\dd z
\right).
\label{eq:S_piecewise_lower}
\end{equation}
The upper boundary layer satisfies the identical estimate. Using
\begin{equation}
\avg{\left(\partial_z\we\right)^2}\le
\frac{1}{4}\avg{|\nabla\ue|^2},
\label{eq:S_piecewise_deriv_bound}
\end{equation}
which follows from incompressibility and the no-slip boundary conditions
\citep{DC1996}, we obtain
\begin{equation}
\left|
\avg{\left(a-\frac{\dd\tau}{\dd z}\right)\we\theta}
\right|
\le
\frac{(1+a)\delta}{16\sqrt{a}}
\left(
a\avg{|\nabla\ue|^2}
+\avg{|\nabla\theta|^2}
\right).
\label{eq:S_piecewise_final}
\end{equation}

Combining inequality~\eqref{eq:S_piecewise_final} with the neutral quotient in equation~\eqref{eq:Ra_min_main} gives the lower bound
\begin{equation}
\sqrt{Ra_{\min}}\ge
\frac{16\sqrt{a}}{(1+a)\delta}.
\label{eq:S_piecewise_Ra}
\end{equation}
For $\delta\ll1$, the wall gradient is
\begin{equation}
Nu=
-\left.\frac{\dd\tau}{\dd z}\right|_{z=0}
=
\frac{1+a(1-2\delta)}{2\delta}
\simeq\frac{1+a}{2\delta}.
\end{equation}
Since $Nu$ decreases with $\delta$, substituting the lower bound on $\delta$ from inequality~\eqref{eq:S_piecewise_Ra} gives
\begin{equation}
Nu\le
\frac{(1+a)^2}{32\sqrt{a}}\,Ra^{1/2}.
\label{eq:S_piecewise_Nu_bound}
\end{equation}

The inequality~\eqref{eq:S_piecewise_Nu_bound} holds for each admissible $a>0$, but $a$ is not a free parameter to be chosen by optimizing this inequality alone. Rather, $a$ is determined self-consistently by the minimum-$Ra$ marginal-mode closure, which, as established in Appendix~\ref{app:background-lagrangian}, is equivalent to minimizing $\mathcal{J}[\tau]=\int_0^1|\dd\tau/\dd z|^2\,\dd z$ subject to the marginal energy-stability constraint. This is distinct from minimizing the prefactor $C_{\mathrm{pw}}(a)=(1+a)^2/(32\sqrt{a})$ in inequality~\eqref{eq:S_piecewise_Nu_bound} with respect to $a$, which would give $a=1/3$. Within the piecewise-profile family, the marginal energy-stability constraint~\eqref{eq:S_piecewise_Ra} fixes $\delta$ in terms of $a$, so minimizing $\mathcal{J}[\tau]$ reduces to minimizing over $a$ alone. For the piecewise profile~\eqref{eq:S_piecewise_tau}, direct computation gives
\begin{equation}
    \mathcal{J}[\tau]
    = \frac{[1+a(1-2\delta)]^2}{2\delta} + a^2(1-2\delta).
    \label{eq:S_piecewise_J}
\end{equation}
For $\delta\ll1$, the dominant term is $(1+a)^2/(2\delta)$. Substituting $\delta=16\sqrt{a}/[(1+a)\sqrt{Ra}]$ from the marginality condition~\eqref{eq:S_piecewise_Ra} yields
\begin{equation}
    \mathcal{J}[\tau]
    \approx \frac{(1+a)^3\sqrt{Ra}}{32\sqrt{a}}.
    \label{eq:S_piecewise_J_asymp}
\end{equation}
Minimizing over $a$ by setting $\dd\mathcal{J}/\dd a=0$ gives $6a=1+a$, hence $a=1/5$, consistent with the numerical result. Substituting into inequality~\eqref{eq:S_piecewise_Nu_bound} gives
\begin{equation}
    Nu \le \frac{9\sqrt{5}}{200}\,Ra^{1/2} \simeq 0.1006\,Ra^{1/2}.
    \label{eq:S_piecewise_final_Nu}
\end{equation}

The piecewise construction is therefore asymptotically informative but quantitatively non-sharp: it yields $\delta=O(Ra^{-1/2})$ and $Nu=O(Ra^{1/2})$, while its coefficient at the asymptotic value $a=1/5$ is approximately $4.1$ times the computed coefficient $0.0245$.

\section{Equivalent background Lagrangian formulation and uniqueness of the optimal profile}\label{app:background-lagrangian}

The direct construction in section~\ref{sec:formulation} starts from the perturbation-energy identity~\eqref{eq:energy_identity}: for a prescribed mean temperature profile $\tau(z)$, one minimizes the neutral Rayleigh number over the unknown balance parameter and the disturbance fields, and then couples the minimizing modes to the exact mean-temperature equation~\eqref{eq:tau}. In this appendix we show that the same closed equations also arise from a Doering--Constantin-type background problem.

For a fixed Rayleigh number $Ra$, define the quadratic energy form
\begin{equation}
    \mathcal{Q}_{Ra}[\tau,a;X]
    = \avg{
        \frac{a}{\sqrt{Ra}}|\nabla\ue|^2
        +\frac{1}{\sqrt{Ra}}|\nabla\theta|^2
        -\left(a-\frac{\dd\tau}{\dd z}\right)\we\theta
    }, \qquad X=(\ue,\theta).
    \label{eq:A_Q_background}
\end{equation}
The equivalence established in equation~\eqref{eq:minRa_spectral_equivalence}
may be written as
\begin{equation}
    Ra \le Ra_{\min}[\tau]
    \quad\Longleftrightarrow\quad
    \mathcal{Q}_{Ra}[\tau,a;X] \ge 0
    \quad\text{for every }a>0\text{ and every admissible }X.
    \label{eq:A_Q_equivalence}
\end{equation}
Thus, at fixed $Ra$, the admissible mean profiles form the set
\begin{equation}
    \mathcal{C}_{Ra} =
    \left\{
        \tau\in H^1(0,1):
        \tau(0)=1,\ \tau(1)=0,\
        \mathcal{Q}_{Ra}[\tau,a;X]\ge0
        \ \forall\,a>0,\ \forall\,X
    \right\}.
    \label{eq:A_convex_set}
\end{equation}
For every fixed pair $(a,X)$, $\mathcal{Q}_{Ra}$ is affine in $\tau$ because
\begin{equation}
    \mathcal{Q}_{Ra}[\tau,a;X]
    = a\left(\frac{\avg{|\nabla\ue|^2}}{\sqrt{Ra}}-\avg{\we\theta}\right)
    +\frac{\avg{|\nabla\theta|^2}}{\sqrt{Ra}}
    +\avg{\frac{\dd\tau}{\dd z}\we\theta}.
    \label{eq:A_Q_affine}
\end{equation}

For a prescribed temperature profile $\tau(z)$ and a fixed balance parameter $a>0$, $\mathcal{Q}_{Ra}$ is strictly convex in $X$. To examine the convexity, consider an arbitrary variation $\delta X=(\delta\ue,\delta\theta)$. The second variation of $\mathcal{Q}_{Ra}$ is
\begin{align}
    \delta^2\mathcal{Q}_{Ra}
    &= \frac{2a}{\sqrt{Ra}}\avg{|\nabla\delta\ue|^2}
      +\frac{2}{\sqrt{Ra}}\avg{|\nabla\delta\theta|^2}
      -2\avg{\left(a-\frac{\dd\tau}{\dd z}\right)\delta\we\,\delta\theta}.
    \label{eq:A_second_variation}
\end{align}
The coupling term can be bounded by the Cauchy--Schwarz and Poincar\'{e} inequalities. Since $\delta\ue$ and $\delta\theta$ satisfy homogeneous boundary conditions,
\begin{equation}
    \|\delta\we\|_2 \le C_P\|\nabla\delta\ue\|_2,
    \qquad
    \|\delta\theta\|_2 \le C_P\|\nabla\delta\theta\|_2.
\end{equation}
Therefore,
\begin{equation}
    \left|
        \avg{\left(a-\frac{\dd\tau}{\dd z}\right)\delta\we\,\delta\theta}
    \right|
    \le C_P^2
    \left\|a-\frac{\dd\tau}{\dd z}\right\|_{\infty}
    \|\nabla\delta\ue\|_2\|\nabla\delta\theta\|_2.
    \label{eq:A_coupling_bound}
\end{equation}
Applying Young's inequality gives
\begin{equation}
    \delta^2\mathcal{Q}_{Ra}
    \ge
    \left(\frac{2a}{\sqrt{Ra}}-\epsilon C_\tau\right)\|\nabla\delta\ue\|_2^2
    +\left(\frac{2}{\sqrt{Ra}}-\frac{C_\tau}{\epsilon}\right)\|\nabla\delta\theta\|_2^2,
    \label{eq:A_convexity_bound}
\end{equation}
where $C_\tau = 2C_P^2\|a-\dd\tau/\dd z\|_{\infty}$. For a given $(\tau,a)$, there exists $\epsilon>0$ such that both coefficients in~\eqref{eq:A_convexity_bound} are positive, provided that the Rayleigh number is below the marginal energy-stability threshold. Consequently, $\delta^2\mathcal{Q}_{Ra}>0$ for every $\delta X\ne\bm{0}$, and $\mathcal{Q}_{Ra}$ is strictly convex in $X$. At the marginal energy-stability point, the smallest eigenvalue of the Hessian vanishes and the quadratic form becomes positive semidefinite, $\mathcal{Q}_{Ra}(X)\ge0$, with equality only for the neutral energy mode.

We now consider the auxiliary problem
\begin{equation}
    \min_{\tau\in\mathcal{C}_{Ra}}
    \mathcal{J}[\tau],
    \qquad
    \mathcal{J}[\tau] = \int_0^1\left|\frac{\dd\tau}{\dd z}\right|^2\dd z.
    \label{eq:A_background_primal}
\end{equation}
The objective $\mathcal{J}$ is strictly convex on the affine space of profiles with fixed boundary values. Indeed, for distinct $\tau_1$ and $\tau_2$ and
$0<s<1$,
\begin{equation}
    \mathcal{J}[s\tau_1+(1-s)\tau_2]
    = s\mathcal{J}[\tau_1]+(1-s)\mathcal{J}[\tau_2]
      -s(1-s)\int_0^1\left|
          \frac{\dd\tau_1}{\dd z}-\frac{\dd\tau_2}{\dd z}
      \right|^2\dd z.
    \label{eq:A_strict_convexity}
\end{equation}
Hence, whenever the feasible set is non-empty and a minimizer exists, the optimal mean profile is unique. This is the same convex structure used in the Doering--Constantin background method \citep{DC1996,DingKerswell20}.

We minimize $\mathcal{J}[\tau]$ subject to the quadratic constraint $\mathcal{Q}_{Ra}\ge0$ and the incompressibility condition by introducing the Lagrangian functional
\begin{align}
    \mathcal{L}
    :={}& \avg{\left(\frac{\dd\tau}{\dd z}\right)^2}
        + 2\sqrt{Ra}\avg{P(\bm{x})\,\nabla\cdot\ue}
    \nonumber\\
    &- 2\sqrt{Ra}\avg{
        \frac{a}{\sqrt{Ra}}|\nabla\ue|^2
        +\frac{1}{\sqrt{Ra}}|\nabla\theta|^2
        -\left(a-\frac{\dd\tau}{\dd z}\right)\we\theta
    }.
    \label{eq:A_displayed_lagrangian}
\end{align}
Similar constrained minimization of $\mathcal{J}[\tau]$ subject to a marginal energy-stability spectral constraint was first employed by \citet{Chini2011} to construct low-dimensional models of porous medium convection. Note that $\mathcal{L}$ is a convex-concave saddle functional of the same type as the background method \citep{DC1996}; its stationarity conditions govern all active marginal modes, and uniqueness of the global optimal solution follows from the strict convexity established above.

Taking first variations of $\mathcal{L}$ with respect to $P$, $\ue$, $\theta$, $a$, and $\tau$ yields the Euler--Lagrange system
\begin{subequations}
\label{eq:A_EL}
\begin{align}
    \nabla\cdot\ue &= 0,
    \label{eq:A_EL_incomp}\\
    \frac{2a}{\sqrt{Ra}}\nabla^2\ue
        +\left(a-\frac{\dd\tau}{\dd z}\right)\theta\bm{e}_z
        -\nabla P &= \bm{0},
    \label{eq:A_EL_momentum}\\
    \frac{2}{\sqrt{Ra}}\nabla^2\theta
        +\left(a-\frac{\dd\tau}{\dd z}\right)\we &= 0,
    \label{eq:A_EL_temperature}\\
    \avg{|\nabla\ue|^2}-\sqrt{Ra}\avg{\we\theta} &= 0,
    \label{eq:A_EL_a}\\
    \frac{\dd^2\tau}{\dd z^2}
        -\sqrt{Ra}\,\frac{\dd \havg{\we\theta}}{\dd z} &= 0.
    \label{eq:A_EL_tau}
\end{align}
\end{subequations}
These are equivalent to equations~\eqref{eq:el} and the mean-temperature equation~\eqref{eq:tau} (the momentum and temperature equations differ by an overall sign, which does not affect the solution). Thus the Lagrangian~\eqref{eq:A_displayed_lagrangian} and the direct construction produce the same governing equations. Strict convexity of $\mathcal{J}$ on $\mathcal{C}_{Ra}$ establishes uniqueness of $\tau$ and hence of the associated Nusselt number $Nu=-\left.\frac{\dd\tau}{\dd z}\right|_{z=0}$. Therefore, equations~\eqref{eq:el} together with the mean-temperature equation~\eqref{eq:tau} admit a unique globally optimal solution.

\bibliographystyle{jfm}

\bibliography{myrefs}

\begin{thebibliography}{46}
\expandafter\ifx\csname natexlab\endcsname\relax\def\natexlab#1{#1}\fi
\def\au#1{#1} \def\ed#1{#1} \def\yr#1{#1}\def\at#1{#1}\def\jt#1{\textit{#1}} \def\bt#1{#1}\def\bvol#1{\textbf{#1}} \def\vol#1{#1} \def\pg#1{#1} \def\publ#1{#1}\def\arxiv#1{#1}\def\org#1{#1}\def\st#1{\textit{#1}}

\bibitem[Ahlers {\em et~al.\/}(2009)Ahlers, Grossmann \& Lohse]{Ahlers2009}
{\sc \au{Ahlers, G.}, \au{Grossmann, S.} \& \au{Lohse, D.}} \yr{2009}  \at{{Heat transfer and large scale dynamics in turbulent Rayleigh--B{\'{e}}nard convection}}.  \jt{Rev. Mod. Phys.}  \bvol{81},  \pg{503--537}.

\bibitem[Alben(2017)]{Alben2017}
{\sc \au{Alben, S.}} \yr{2017}  \at{Optimal convection cooling flows in general 2d geometries}.  \jt{J. Fluid Mech.}  \bvol{814},  \pg{484–509}.

\bibitem[Bouillaut {\em et~al.\/}(2019)Bouillaut, Lepot, Aumaitre \& Gallet]{Bouillaut}
{\sc \au{Bouillaut, V.}, \au{Lepot, S.}, \au{Aumaitre, S.} \& \au{Gallet, B.}} \yr{2019}  \at{Transition to the ultimate regime in a radiatively driven convection experiment}.  \jt{J. Fluid Mech.}  \bvol{861},  \pg{R5}.

\bibitem[Burns {\em et~al.\/}(2020)Burns, Vasil, Oishi, Lecoanet \& Brown]{Burns2020}
{\sc \au{Burns, K.~J.}, \au{Vasil, G.~M.}, \au{Oishi, J.~S.}, \au{Lecoanet, D.} \& \au{Brown, B.~P.}} \yr{2020}  \at{{Dedalus: A flexible framework for numerical simulations with spectral methods}}.  \jt{Phys. Rev. Research}  \bvol{2}~(2),  \pg{023068}.

\bibitem[Busse(1969)]{Busse1969}
{\sc \au{Busse, F.~H.}} \yr{1969}  \at{On {H}oward's upper bound for heat transport by turbulent convection}.  \jt{J. Fluid Mech.}  \bvol{37}~(3),  \pg{457--477}.

\bibitem[Chavanne {\em et~al.\/}(1997)Chavanne, Chilla, Castaing, Hebral, Chabaud \& Chaussy]{Chavanne97}
{\sc \au{Chavanne, X.}, \au{Chilla, F.}, \au{Castaing, B.}, \au{Hebral, B.}, \au{Chabaud, B.} \& \au{Chaussy, J.}} \yr{1997}  \at{{Observation of the ultimate regime in Rayleigh--B\'enard convection}}.  \jt{Phys. Rev. Lett.}  \bvol{79},  \pg{3648--3651}.

\bibitem[Chill{\`{a}} \& Schumacher(2012)]{Chilla2012}
{\sc \au{Chill{\`{a}}, F.} \& \au{Schumacher, J.}} \yr{2012}  \at{{New perspectives in turbulent Rayleigh--B{\'{e}}nard convection}}.  \jt{The Eu. Phys. J. E}  \bvol{35},  \pg{58}.

\bibitem[Chini {\em et~al.\/}(2011)Chini, Dianati, Zhang \& Doering]{Chini2011}
{\sc \au{Chini, G.~P.}, \au{Dianati, N.}, \au{Zhang, Z.} \& \au{Doering, C.~R.}} \yr{2011}  \at{Low-dimensional models from upper bound theory}.  \jt{Phys. D: Nonlinear Phenom.}  \bvol{240}~(2),  \pg{241--248}.

\bibitem[Ding(2024)]{Ding2024}
{\sc \au{Ding, Z.}} \yr{2024}  \at{Bounding heat transport in supergravitational turbulent thermal convection}.  \jt{J. Fluid Mech.}  \bvol{1001},  \pg{A56}.

\bibitem[Ding \& Kerswell(2020)]{DingKerswell20}
{\sc \au{Ding, Z.} \& \au{Kerswell, R.~R.}} \yr{2020}  \at{{Exhausting the background approach for bounding the heat transport in Rayleigh-Benard convection}}.  \jt{J. Fluid Mech.}  \bvol{889},  \pg{A33}.

\bibitem[Ding \& Ouyang(2023)]{Ding2023}
{\sc \au{Ding, Z.} \& \au{Ouyang, Z.}} \yr{2023}  \at{Penetrative convection: heat transport with marginal stability assumption}.  \jt{J. Fluid Mech.}  \bvol{960},  \pg{A26}.

\bibitem[Ding \& Wu(2021)]{Ding2021}
{\sc \au{Ding, Z.} \& \au{Wu, J.}} \yr{2021}  \at{Coherent heat transport in two-dimensional penetrative rayleigh–bénard convection}.  \jt{J. Fluid Mech.}  \bvol{920},  \pg{A48}.

\bibitem[Doering \& Constantin(1996)]{DC1996}
{\sc \au{Doering, C.~R.} \& \au{Constantin, P.}} \yr{1996}  \at{Variational bounds on energy dissipation in incompressible flows. iii. convection}.  \jt{Phys. Rev. E}  \bvol{53},  \pg{5957--5981}.

\bibitem[Doering \& Tobasco(2019)]{Doering2019}
{\sc \au{Doering, C.~R.} \& \au{Tobasco, I.}} \yr{2019}  \at{On the optimal design of wall-to-wall heat transport}.  \jt{Commun. Pure Appl. Math.}  \bvol{72}~(11),  \pg{2385--2448}.

\bibitem[Hassanzadeh {\em et~al.\/}(2014)Hassanzadeh, Chini \& Doering]{Hassanzadeh2014}
{\sc \au{Hassanzadeh, P.}, \au{Chini, G.~P.} \& \au{Doering, C.~R.}} \yr{2014}  \at{Wall to wall optimal transport}.  \jt{J. Fluid Mech.}  \bvol{751},  \pg{627--662}.

\bibitem[He {\em et~al.\/}(2026)He, Motoki, Deguchi \& Kawahara]{He2026}
{\sc \au{He, X.}, \au{Motoki, S.}, \au{Deguchi, K.} \& \au{Kawahara, G.}} \yr{2026}  \at{High-{R}ayleigh-number asymptotic classical scaling in three-dimensional steady natural convection}.  \jt{J. Fluid Mech.}  \bvol{1028},  \pg{A5}.

\bibitem[Howard(1963)]{Howard1963}
{\sc \au{Howard, L.~N.}} \yr{1963}  \at{Heat transport by turbulent convection}.  \jt{J. Fluid Mech.}  \bvol{17},  \pg{405--432}.

\bibitem[Howard(1972)]{Howard1972}
{\sc \au{Howard, L.~N.}} \yr{1972}  \at{Bounds on flow quantities}.  \jt{Annu. Rev. Fluid Mech.}  \bvol{4}~(1),  \pg{473--494}.

\bibitem[Iyer {\em et~al.\/}(2020)Iyer, Scheel, Schumacher \& Sreenivasan]{Iyer}
{\sc \au{Iyer, K.}, \au{Scheel, J.}, \au{Schumacher, J.} \& \au{Sreenivasan, K.}} \yr{2020}  \at{{Classical 1/3 scaling of convection holds up to Ra = 10$^{15}$}}.  \jt{Proc. Natl. Acad. Sci. USA}  \bvol{117},  \pg{7594--7598}.

\bibitem[Jiang {\em et~al.\/}(2022)Jiang, Wang, Liu \& Sun]{Jiang}
{\sc \au{Jiang, H.}, \au{Wang, D.}, \au{Liu, S.} \& \au{Sun, C.}} \yr{2022}  \at{Experimental evidence for the existence of the ultimate regime in rapidly rotating turbulent thermal convection}.  \jt{Phys. Rev. Lett.}  \bvol{129},  \pg{204502}.

\bibitem[Kooloth {\em et~al.\/}(2021)Kooloth, Sondak \& Smith]{Kooloth2021}
{\sc \au{Kooloth, P.}, \au{Sondak, D.} \& \au{Smith, L.~M.}} \yr{2021}  \at{{Coherent solutions and transition to turbulence in two-dimensional Rayleigh--B\'enard convection}}.  \jt{Phys. Rev. Fluids}  \bvol{6},  \pg{013501}.

\bibitem[Kraichnan(1962)]{Kraichnan1962}
{\sc \au{Kraichnan, R.~H.}} \yr{1962}  \at{{Turbulent thernal convection at arbitrary Prandtl number}}.  \jt{Phys. Fluids}  \bvol{5},  \pg{1374--1389}.

\bibitem[Lepot {\em et~al.\/}(2018)Lepot, Aumaitre \& Gallet]{Lepot}
{\sc \au{Lepot, S.}, \au{Aumaitre, S.} \& \au{Gallet, B.}} \yr{2018}  \at{Radiative heating achieves the ultimate regime of thermal convection}.  \jt{Proc. Natl. Acad. Sci. USA}  \bvol{115},  \pg{8937}.

\bibitem[Malkus(1954)]{Malkus1954}
{\sc \au{Malkus, W. V.~R.}} \yr{1954}  \at{{The heat transport and spectrum of thermal turbulence}}.  \jt{Proc. Roy. Soc.}  \bvol{A225},  \pg{196--212}.

\bibitem[Motoki {\em et~al.\/}(2018)Motoki, Kawahara \& Shimizu]{Motoki2018}
{\sc \au{Motoki, S.}, \au{Kawahara, G.} \& \au{Shimizu, M.}} \yr{2018}  \at{{Maximal heat transfer between two parallel plates}}.  \jt{J. Fluid Mech.}  \bvol{851},  \pg{R4}.

\bibitem[Motoki {\em et~al.\/}(2021)Motoki, Kawahara \& Shimizu]{Motoki2021}
{\sc \au{Motoki, S.}, \au{Kawahara, G.} \& \au{Shimizu, M.}} \yr{2021}  \at{{Multi-scale steady solution for Rayleigh--B{\'e}nard convection}}.  \jt{J. Fluid Mech.}  \bvol{914},  \pg{A14}.

\bibitem[Niemela {\em et~al.\/}(2000)Niemela, Skrbek, Sreenivasan \& Donnelly]{Niemela}
{\sc \au{Niemela, J.}, \au{Skrbek, L.}, \au{Sreenivasan, K.} \& \au{Donnelly, R.}} \yr{2000}  \at{Turbulent convection at very high rayleigh numbers}.  \jt{Nature}  \bvol{404},  \pg{837}.

\bibitem[O'Connor {\em et~al.\/}(2021)O'Connor, Lecoanet \& Anders]{OConnor2021}
{\sc \au{O'Connor, L.}, \au{Lecoanet, D.} \& \au{Anders, E.~H.}} \yr{2021}  \at{{Marginally-stable thermal equilibria of Rayleigh-B\'{e}nard convection}}.  \jt{Phys. Rev. Fluids}  \bvol{6},  \pg{093501}.

\bibitem[Ouyang {\em et~al.\/}(2025)Ouyang, Wang, Li, Wen \& Ding]{Ouyang2025}
{\sc \au{Ouyang, Z.}, \au{Wang, Q.}, \au{Li, K.}, \au{Wen, B.} \& \au{Ding, Z.}} \yr{2025}  \at{Touching the classical scaling in penetrative convection}.  \jt{Proc. Natl. Acad. Sci. USA}  \bvol{122},  \pg{e2418468122}.

\bibitem[Plasting \& Kerswell(2003)]{Plasting2003}
{\sc \au{Plasting, S.~C.} \& \au{Kerswell, R.~R.}} \yr{2003}  \at{{Improved upper bound on the energy dissipation rate in plane Couette flow: the full solution to Busse's problem and the Constantin--Doering--Hopf problem with one-dimensional background field}}.  \jt{J. Fluid Mech.}  \bvol{477},  \pg{363--379}.

\bibitem[Potticary {\em et~al.\/}(2021)Potticary, Hall, Guo, Price \& Hall]{Potticary2021}
{\sc \au{Potticary, J.}, \au{Hall, C.~L.}, \au{Guo, R.}, \au{Price, S.~L.} \& \au{Hall, S.~R.}} \yr{2021}  \at{On the application of strong magnetic fields during organic crystal growth}.  \jt{Cryst. Growth Des.}  \bvol{21}~(11),  \pg{6254--6265}.

\bibitem[Qu \& Mudawar(2002)]{Qu2002}
{\sc \au{Qu, Weilin} \& \au{Mudawar, Issam}} \yr{2002}  \at{Analysis of three-dimensional heat transfer in micro-channel heat sinks}.  \jt{Int. J Heat Mass trans.}  \bvol{45}~(19),  \pg{3973--3985}.

\bibitem[Sondak {\em et~al.\/}(2015)Sondak, Smith \& Waleffe]{Sondak2015}
{\sc \au{Sondak, D.}, \au{Smith, L.~M.} \& \au{Waleffe, F.}} \yr{2015}  \at{{Optimal heat transport solutions for Rayleigh--B{\'{e}}nard convection}}.  \jt{J. Fluid Mech.}  \bvol{784},  \pg{565--595}.

\bibitem[Souza {\em et~al.\/}(2020)Souza, Tobasco \& Doering]{Souza2020}
{\sc \au{Souza, A.~N.}, \au{Tobasco, I.} \& \au{Doering, C.~R.}} \yr{2020}  \at{Wall-to-wall optimal transport in two dimensions}.  \jt{J. Fluid Mech.}  \bvol{889},  \pg{A34}.

\bibitem[Spiegel(1963)]{Spiegel1963}
{\sc \au{Spiegel, E.~A.}} \yr{1963}  \at{{A generalization of mixing-length theory of turbulent convection}}.  \jt{Astrophys. J.}  \bvol{138},  \pg{216--225}.

\bibitem[Tobasco \& Doering(2017)]{Tobasco2017}
{\sc \au{Tobasco, I.} \& \au{Doering, C.~R.}} \yr{2017}  \at{Optimal wall-to-wall transport by incompressible flows}.  \jt{Phys. Rev. Lett.}  \bvol{118},  \pg{264502}.

\bibitem[Urban {\em et~al.\/}(2011)Urban, Musilov{\'a} \& Skrbek]{Urban2011}
{\sc \au{Urban, P.}, \au{Musilov{\'a}, V.} \& \au{Skrbek, L.}} \yr{2011}  \at{{Efficiency of heat transfer in turbulent Rayleigh-B{\'e}nard convection}}.  \jt{Phys. Rev. Lett.}  \bvol{107}~(1),  \pg{014302}.

\bibitem[Vegad \& Bhatt(2014)]{Vegad2014}
{\sc \au{Vegad, M.} \& \au{Bhatt, N.~M.}} \yr{2014}  \at{{Review of some aspects of single crystal growth using Czochralski crystal growth technique}}.  \jt{Procedia Technol.}  \bvol{14},  \pg{438--446}.

\bibitem[Waleffe {\em et~al.\/}(2015)Waleffe, Boonkasame \& Smith]{Waleffe2015}
{\sc \au{Waleffe, F.}, \au{Boonkasame, A.} \& \au{Smith, L.~M.}} \yr{2015}  \at{{Heat transport by coherent Rayleigh--B{\'{e}}nard convection}}.  \jt{Phys. Fluids}  \bvol{27},  \pg{051702}.

\bibitem[Wen {\em et~al.\/}(2022{\natexlab{{\em a\/}}})Wen, Ding, Chini \& Kerswell]{Wen2022RSTA}
{\sc \au{Wen, B.}, \au{Ding, Z.}, \au{Chini, G.} \& \au{Kerswell, R.}} \yr{2022{\natexlab{{\em a\/}}}}  \at{{Heat transport in Rayleigh--B{\'e}nard convection with linear marginality}}.  \jt{Phil. Trans. R. Soc. A}  \bvol{380}~(2225).

\bibitem[Wen {\em et~al.\/}(2022{\natexlab{{\em b\/}}})Wen, Goluskin \& Doering]{Wen2022JFM}
{\sc \au{Wen, B.}, \au{Goluskin, D.} \& \au{Doering, C.~R.}} \yr{2022{\natexlab{{\em b\/}}}}  \at{{Steady Rayleigh--B{\'{e}}nard convection between no-slip boundaries}}.  \jt{J. Fluid Mech.}  \bvol{933},  \pg{R4}.

\bibitem[Wen {\em et~al.\/}(2020)Wen, Goluskin, LeDuc, Chini \& Doering]{Wen2020JFM}
{\sc \au{Wen, B.}, \au{Goluskin, D.}, \au{LeDuc, M.}, \au{Chini, G.~P.} \& \au{Doering, C.~R.}} \yr{2020}  \at{{Steady Rayleigh--B{\'{e}}nard convection between stress-free boundaries}}.  \jt{J. Fluid Mech.}  \bvol{905},  \pg{R4}.

\bibitem[Whitehead \& Doering(2011)]{Whitehead}
{\sc \au{Whitehead, J.} \& \au{Doering, C.}} \yr{2011}  \at{Ultimate state of two-dimensional rayleigh-b\'enard convection between free-slip fixed-temperature boundaries}.  \jt{Phys. Rev. Lett.}  \bvol{106},  \pg{244501}.

\bibitem[Xia(2013)]{Xia2013}
{\sc \au{Xia, K-.Q}} \yr{2013}  \at{Current trends and future directions in turbulent thermal convection}.  \jt{Theor. Appl. Mech. Lett.}  \bvol{3},  \pg{052001}.

\bibitem[Xia {\em et~al.\/}(2025)Xia, Chong, Ding \& Zhang]{Xia2025}
{\sc \au{Xia, K-.Q}, \au{Chong, K.}, \au{Ding, G.} \& \au{Zhang, L.}} \yr{2025}  \at{Some fundamental issues in buoyancy-driven flows with implications for geophysical and astrophysical systems}.  \jt{Acta Mech. Sinica}  \bvol{41},  \pg{324287}.

\bibitem[Yaron \& Gadi(2019)]{Shenhav2019Patent}
{\sc \au{Yaron, S.} \& \au{Gadi, G.}} \yr{2019} Cooling with anti-stokes fluorescence. US Patent App. 16/318,137 (Publication No. US20190154316A1).

\end{thebibliography}

\end{document}